\documentclass[usenatbib]{mn2e}
\usepackage{times,graphicx,url}

\newcommand{\be}{\begin{equation}}
\newcommand{\ee}{\end{equation}}

\def\lesssim{\mathrel{\hbox{\rlap{\hbox{\lower4pt\hbox{$\sim$}}}\hbox{$<$}}}}
\def\gtrsim{\mathrel{\hbox{\rlap{\hbox{\lower4pt\hbox{$\sim$}}}\hbox{$>$}}}}
\newcommand{\pivec}{\mbox{\boldmath $\pi$}}

\def\apj{ApJ}

\def\aap{A\&\hskip-1pt A}

\def\mnras{MNRAS}
\def\pasp{PASP}

\title[OGLE-2009-BLG-023/MOA-2009-BLG-028]{OGLE-2009-BLG-023/MOA-2009-BLG-028: 
Characterization of a Binary Microlensing Event Based on Survey Data}

\author[K.-H. Hwang, et al.]{K.-H. Hwang$^{1}$, C. Han$^{1,16}$, A.\ Udalski$^{2}$, T.\ Sumi$^{3}$, A.\ Gould$^{4}$, \newauthor
and \newauthor
M. Jaroszy\'nski$^{2}$, M. Kubiak$^{2}$, M.K. Szyma\'nski$^{2}$, G.\ Pietrzy\'nski$^{2,5}$, I. Soszy\'nski$^{2}$, \newauthor
O. Szewczyk$^{2,5}$, K.\ Ulaczyk$^{2}$, \'L. Wyrzykowski$^{2,6}$ \newauthor
(The OGLE Collaboration), \newauthor
F. Abe$^{3}$, D.P.\ Bennett$^{7}$, I.A.\ Bond$^{8}$, C.S.\ Botzler$^{9}$, M.\ Freeman$^{9}$, A. Fukui$^{3}$, \newauthor
K.\ Furusawa$^{3}$, J.B.\ Hearnshaw$^{10}$, Y.\ Itow$^{3}$, K.\ Kamiya$^{3}$, P.M.\ Kilmartin$^{11}$, \newauthor
A.\ Korpela$^{12}$, W. Lin$^{8}$, C.H.\ Ling$^{8}$, K.\ Masuda$^{3}$, Y.\ Matsubara$^{3}$, N.\ Miyake$^{3}$, \newauthor
Y.\ Muraki$^{13}$, K.\ Ohnishi$^{14}$, Y.C. Perrott$^{6}$, N.J.\ Rattenbury$^{9}$, To.\ Saito$^{15}$, T.\ Sako$^{3}$, \newauthor
D.J.\ Sullivan$^{12}$, W.L.\ Sweatman$^{8}$, P.J.\ Tristram$^{12}$, P.C.M. Yock$^{9}$ \newauthor
(The MOA Collaboration)\\
$^{1}$Department of Physics, Chungbuk National University, Cheongju 361-763, Republic of Korea\\
$^{2}$Warsaw University Observatory, Al. Ujazdowskie 4, 00-478 Warszawa, Poland\\
$^{3}$Solar-Terrestrial Environment Laboratory, Nagoya University, Nagoya, 464-8601, Japan\\
$^{4}$Department of Astronomy, Ohio State University, 140 W. 18th Ave., Columbus, OH 43210, USA\\
$^{5}$Universidad de Concepci\'on, Departamento de Fisica, Casilla 160-C, Concepci\'on, Chile\\
$^{6}$Institute of Astronomy, University of Cambridge, Madingley Road, Cambridge CB3 0HA, UK\\
$^{7}$Department of Physics, University of Notre Dame, Notre Dame, IN 46556, USA\\
$^{8}$Institute of Information and Mathematical Sciences, Massey University, Private Bag 102-904, North Shore Mail Centre, Auckland, New Zealand\\
$^{9}$Department of Physics, University of Auckland, Private Bag 92019, Auckland, New Zealand\\
$^{10}$University of Canterbury, Department of Physics and Astronomy, Private Bag 4800, Christchurch 8020, New Zealand\\
$^{11}$Mt. John Observatory, P.O. Box 56, Lake Tekapo 8770, New Zealand\\
$^{12}$School of Chemical and Physical Sciences, Victoria University, Wellington, New Zealand\\
$^{13}$Department of Physics, Konan University, Nishiokamoto 8-9-1, Kobe 658-8501, Japan\\
$^{14}$Nagano National College of Technology, Nagano 381-8550, Japan\\
$^{15}$Tokyo Metropolitan College of Industrial Technology, Tokyo 116-8523, Japan\\
$^{16}$Corresponding author}

\begin{document}

\maketitle

\begin{abstract}
We report the result of the analysis of the light curve of a 
caustic-crossing binary-lens microlensing event 
OGLE-2009-BLG-023/MOA-2009-BLG-028.  Even though the event was 
observed solely by survey experiments, we could uniquely determine 
the mass of the lens and distance to it by simultaneously measuring 
the Einstein radius and lens parallax.  From this, we find that the 
lens system is composed of M-type dwarfs with masses $(0.50\pm 0.07)
\ M_\odot$ and $(0.15\pm 0.02)\ M_\odot$ located in the Galactic disk 
with a distance of $\sim 1.8$ kpc toward the Galactic bulge direction.  
The event demonstrates that physical lens parameters of binary-lens 
events can be routinely determined from future high-cadence lensing 
surveys and thus microlensing can provide a new way to study Galactic 
binaries.
\end{abstract}

\begin{keywords}
gravitational lensing: micro -- binaries: general
\end{keywords}

\section{Introduction}

Microlensing occurs when a foreground star (lens) is closely aligned to 
a background star (source) and the light from the source is refracted by 
the gravity of the lens.  The phenomenon causes splits and distortions
of the background stellar image.  For source stars in the Galaxy, the 
separation between the split images is of an order of milli-arcsec and 
thus cannot be directly observed.  However, the phenomenon can be observed 
through the brightness change of the source star caused by the change of 
the relative lens-source separation \citep{paczynski86}.  Currently, two 
groups \citep{udalski03,bond02} are conducting survey observations to 
detect microlensing events by observing stars located toward the Galactic 
bulge direction.  From these surveys, more than 500 events are being 
detected every year.

For most cases of Galactic microlensing events where source stars are 
lensed by a single foreground star, the lensing light curve is represented 
by
\begin{equation}
A(u)={u^2+2\over u(u^2+4)^{1/2}},\qquad
u(t)=\left[ \left( {t-t_0\over t_{\rm E}}\right)^2+u_0^2\right]^{1/2}.
\label{eq1}
\end{equation}
Here $u$ represents the lens-source separation normalized by the Einstein 
radius $\theta_{\rm E}$, $t_{\rm E}$ is the time required for the source 
to transit the Einstein radius (Einstein time scale), $t_0$ is the time 
of the closest lens-source approach, and $u_0$ is the lens-source separation 
at that moment.  Among the lensing parameters of $t_{\rm E}$, $u_0$ and $t_0$
characterizing lensing light curves, only the Einstein time scale provides 
information about the lens because it is related to the physical parameters 
of the lens mass $M$, relative lens-source parallax $\pi_{\rm rel}={\rm AU}
(D_{\rm L}^{-1}-D_{\rm S}^{-1})$, and proper motion $\mu$ by
\begin{equation}
t_{\rm E}={\theta_{\rm E}\over \mu};\qquad
\theta_{\rm E}=(\kappa M\pi_{\rm rel})^{1/2},
\label{eq2}
\end{equation}
where $\kappa=4G/(c^2{\rm AU})$ and $D_{\rm L}$ and $D_{\rm S}$ represent
the distances to the lens and source star, respectively.  However, the 
time scale results from the combination of the underlying physical lens 
parameters. As a result, it is difficult to uniquely determine the 
individual lens parameters from the time scale alone.

The degeneracy of the lens parameters can be partly lifted by measuring 
either a proper motion or a lens parallax, and can be completely broken 
by measuring both.  The proper motion is related to the Einstein radius 
and the time scale by $\mu=\theta_{\rm E}/t_{\rm E}$ and thus measuring 
$\mu$ is equivalent to measuring $\theta_{\rm E}$.  Einstein radii are 
generally measured from the deviation of the light curve from that of 
a point-source event caused by the finite-source effect \citep{nemiroff94, 
witt94, gould94}.  The microlens parallax is defined by the ratio of the 
Earth's orbit to the Einstein radius projected on the observer plane, 
$\tilde{r}_{\rm E}$, i.e., 
\begin{equation}
\pi_{\rm E}={{\rm AU}\over \tilde{r}_{\rm E}}.
\label{eq3}
\end{equation}
Lens parallaxes are generally measured by analyzing the deviation of
the light curve caused by the change of the observer's position over 
the course of the event due to the orbital motion of the Earth around 
the Sun \citep{refsdal66, gould92, smith03}.  As a result, parallaxes 
are usually measured for events with long time scales that are comparable 
to a significant portion of the Earth's orbital period, i.e.\ 1 yr.  
If both Einstein radius and lens parallax are measured, the mass and 
the distance to the lens are uniquely determined by
\begin{equation}
M={\theta_{\rm E}\over \kappa \pi_{\rm E}},\qquad
D_{\rm L}={{\rm AU}\over \pi_{\rm E}\theta_{\rm E}+\pi_{\rm S}},
\label{eq4}
\end{equation}
respectively.  Here $\pi_{\rm S}={\rm AU}/D_{\rm S}$ represents the 
parallax of the source star.

A fraction of lensing events are produced by lenses composed of binary 
masses \citep{mao91}.  These binary-lens events provide a good chance 
to determine the physical parameters of lenses.  The first reason for 
this is that most binary-lens events are detected through the channel 
of caustic crossings.  The magnification gradient in the region around 
the caustic is so steep that finite-source effect is always manifested 
in the light curves of caustic-crossing events.  If the caustic-crossing 
part of the light curve is resolved, then, it is possible to measure the 
Einstein radius.  Second, the average mass of binary lenses is heavier 
than that of single lenses and thus the time scales of binary events 
tend to be longer than those of single-lens events.  This implies that 
the chance to measure lens parallaxes is higher.

Despite these advantages, mass measurements of binary lenses by measuring 
both $\theta_{\rm E}$ and $\pi_{\rm E}$ were possible only for a handful 
number of events.  These events include 
EROS-2000-BLG-5 \citep{alcock01, an02, gould04},
OGLE-2002-BLG-069 \citep{kubas05}, 
OGLE-2005-BLG-071 \citep{udalski05, dong09},
OGLE-2005-BLG-153 \citep{hwang10b},
OGLE-2006-BLG-109 \citep{gaudi08, bennett10},
MOA-2007-BLG-192 \citep{bennett08}, 
MOA-2009-BLG-016 \citep{hwang10a}, and
OGLE-2009-BLG-092/MOA-2009-BLG-137 \citep{ryu10}.
Among these events, three (OGLE-2005-BLG-071, OGLE-2006-BLG-109, 
MOA-2007-BLG-192) are planetary events for which intensive follow-up 
observations were conducted during the events.  Unfortunately, follow-up 
observations are not being conducted for general binary events due to 
the limited observational resource.

However, the situation is rapidly changing with the instrumental 
upgrade of the survey experiments.  The second phase of the MOA 
survey experiment started in 2006 by replacing its old 0.6 m 
telescope with a new 1.8 m telescope equipped with a camera of 
2.2 ${\rm deg}^2$ field of view.  In 2010, the OGLE group started 
its fourth phase survey observations with a new camera of $1.4\ 
{\rm deg}^2$ field of view that is 3.5 times wider than that of 
the camera used in the third phase experiment.  With the improved 
photometric precision combined with the increased monitoring cadence 
provided by the upgraded instrument, the survey experiments detect 
much more events.  For example, the MOA group detected 563 events 
in 2009 season compared to 56 events in 2005.  In addition, it became 
possible to constrain lenses based on the data obtained from survey 
observations alone for an increasing number of events.

In this paper, we present the result of the analysis of a 
caustic-crossing binary-lens microlensing event 
OGLE-2009-BLG-023/MOA-2009-BLG-028.  This event demonstrates that 
physical parameters of binary lenses can be constrained based on 
the data obtained from survey observations.  We provide the physical 
parameters of the lens system.  We also discuss the use of future 
microlensing surveys in the studies of Galactic binaries.

\begin{figure*}
\begin{center}
\includegraphics[width=0.7\textwidth]{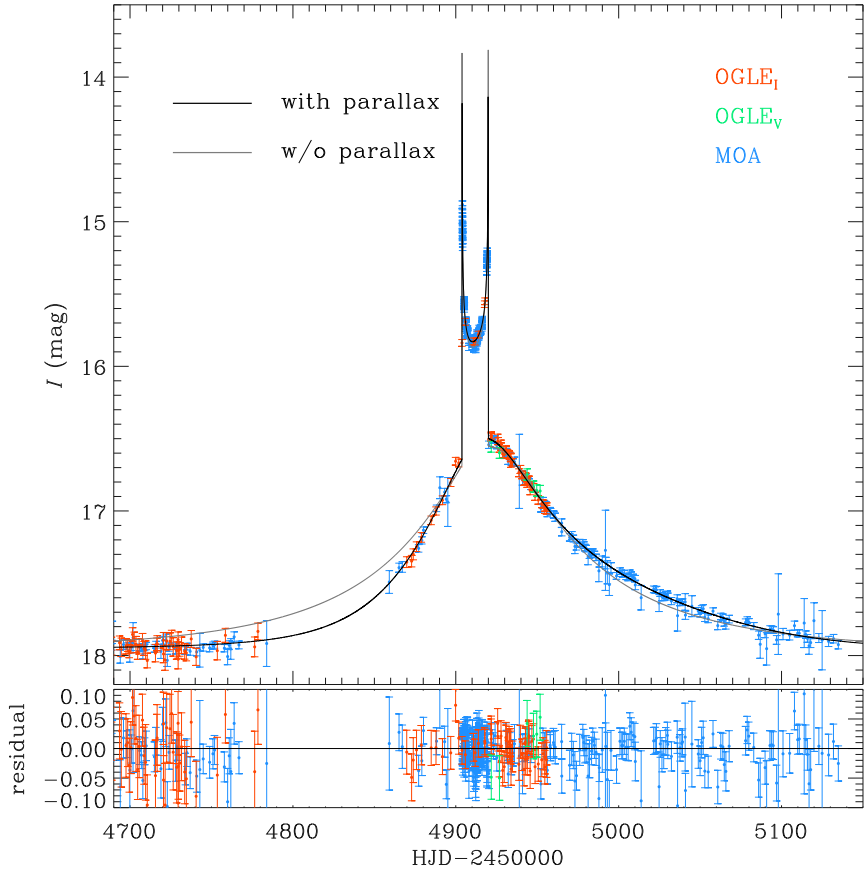}
\end{center}
\caption{
Light curve of the microlensing event OGLE-2009-BLG-023/MOA-2009-BLG-028.
Also presented are the model curves for the best-fit solutions with 
and without the parallax effect.  We note that the MOA data except the 
caustic-induced perturbation part ($2454900 \lesssim {\rm HJD} \lesssim
2454920$) are binned for a clear view.  The zoomed view of the 
caustic-induced part of the light curve is presented in Fig.\ 4.}
\label{fig:one}
\end{figure*}

\section{Observation}

Figure \ref{fig:one} shows the light curve of the event 
OGLE-2009-BLG-023/MOA-2009-BLG-028.  As evidenced by the two 
strong perturbations at ${\rm HJD}\sim 2454905$ and 2454920 and the 
characteristic ``U''-shape trough region between them, the event is 
a typical caustic-crossing binary-lens event.  The brightening of 
the source star was noticed in the early 2009 bulge season by 
both OGLE and MOA survey experiments using the 1.3 m Warsaw telescope 
of Las Campanas Observatory in Chile and the 1.8 m of Mt. John 
Observatory in New Zealand, respectively.  The perturbation produced 
by the first caustic crossing was detected on March 12 by both survey 
experiments.  The time gap during $2454780 \lesssim {\rm HJD}\lesssim 
2454860$ corresponds to the period during which the bulge could not be 
seen.  We note that OGLE observation was stopped on ${\rm HJD}\sim 
2454960$ for the upgrade of the camera.  In addition, no follow-up 
observations of the event was conducted.  Nevertheless, the light 
curve of the event was well covered during the whole 2009 bulge season.

For the analysis of the light curve, we use 1058 $I$-band and 15 
$V$-band OGLE images taken during $2453417 \lesssim {\rm HJD}\lesssim 
2454955$ and $2453470 \lesssim {\rm HJD}\lesssim 2454951$, respectively.  
The MOA data is composed of 1298 $R$-band images taken during $2454495 
\lesssim {\rm HJD}\lesssim 2455134$.  The photometry was processed by 
the individual groups using their own software.

\section{Modeling}

Light curves of binary-lensing events result from a complex phenomenology 
and thus exhibit an astonishing diversity \citep{schneider86}. As a 
result, modeling light curves is a difficult task. One important 
difficulty arises due to the large number of parameters to be included 
in modeling. These parameters are needed to describe various features 
of the light curve. To describe light curves of standard single-lens 
events, a set of three parameters of $t_{\rm E}$, $t_0$, and $u_0$ 
are needed. To describe the deviation caused by the lens binarity, 
an additional set of three binary parameters is needed. These binary 
parameters include the mass ratio between the lens components, $q$, 
the projected binary separation in units of the Einstein radius, $s$, 
and the angle of the source trajectory with respect to the binary axis, 
$\alpha$.  Since OGLE-2009-BLG-023/MOA-2009-BLG-028 is a caustic-crossing 
binary event, an additional parameter of the normalized source radius, 
$\rho_\star\equiv \theta_\star/ \theta_{\rm E}$, is needed to account 
for the finite-source effect.  Here $\theta_\star$ represents the 
angular radius of the source star.  In addition, the event lasted 
throughout the whole 2009 bulge season and thus it is needed to 
check the possibility of deviations induced by the parallax effect. To
incorporate the parallax effect, it is required to include two parallax
parameters $\pi_{{\rm E},N}$ and $\pi_{{\rm E},E}$, which are the
components of the microlens-parallax vector $\pivec_{\rm E}$ projected
on the sky in the north and east celestial coordinates, respectively,
where the direction of the parallax vector is that of the lens-source
relative motion in the frame of the Earth at the peak of the event. 
Due to the shear size of the parameter space, brute-force searches 
for solutions are very difficult and extremely time-consuming.

The second important difficulty in modeling binary-lensing light 
curves is caused by the complexity of $\chi^2$ surface.  This 
complexity implies that even if a solution that apparently describes 
the observed light curve is found, it is difficult to be sure that 
all possible minima have been investigated \citep{dominik99a,dominik99b}.
As a result, a simple downhill approach to search for solutions cannot 
be used.

For efficient modeling but avoiding the difficulties mentioned
above, we use a hybrid approach of parameter searches. In this
approach, grid searches are conducted over the space of a subset of
parameters and the remaining parameters are searched for by letting
them vary so that they result in minimum $\chi^2$ at each set of the 
grid parameters. We choose $s$, $q$, and $\alpha$ as the grid parameters
because they are related to the light curve features in a complex way
such that a small change in the values of the parameters can lead to
dramatic changes in the resulting light curve. On the other hand, the
other parameters are more directly related to the light curve features 
and thus they can be searched for by using a downhill approach.  For 
the $\chi^2$ minimization in the downhill approach, we use a Markov
Chain Monte Carlo method. Once the $\chi^2$ minima of the individual 
grid points are determined, the best-fit model is obtained by comparing 
their $\chi^2$ values.  We investigate the degeneracy of solutions by 
probing local minima that appear in the space of the grid parameters.

The last major difficulty in binary-lensing modeling arises due to 
the fact that the modeling requires large computations. Finding 
the best-fit parameters and evaluating their uncertainties requires
generating a large number of trial model light curves. The problem 
is that most binary-lensing events exhibit deviations induced by the
finite-source effect and calculating finite-source magnifications
requires intensive computations. Formally, the magnification of a finite
source can be calculated by integrating the point-source magnifications 
of the elements of the the source star over its surface.  However, 
this approach is impractical due to the divergent nature of the 
magnification near the caustic and the large amount of computation 
time required for precision calculation. Although there exist
semi-analytic approximations \citep{schneider86}, it is not precise
enough to describe the caustic-crossing feature of lensing light 
curves.  Therefore, an efficient method of finite-magnification 
calculation is important for binary-lensing modeling.

For the finite-source magnification calculation, we use a modified
version of the ray-shooting method. In the usual ray-shooting method,
one shoots a large number of uniformly spaced rays in the image 
plane, and determine which ones land on the surface of the source
using the lens equation.  Then, the magnification corresponding to a
source position is calculated as the ratio between the number density 
of rays on the image plane to that of rays on the source surface
\citep{kayser86, wambsganss90}.  Based on this basic scheme, we 
minimize the computation time by using the following methods.
\begin{enumerate}
\item
Magnification map making\\
Instead of calculating the magnifications corresponding to the 
individual source positions, we construct magnification maps of the 
region encompassing perturbation regions \citep{dong06}. The main 
advantage of this method is that once a map for a fixed ($s$,$q$) 
parameter set is constructed, one can reuse it for the calculations 
of many light curves resulting from different combinations of other 
parameters instead of re-shooting rays all over again.
\item
Minimization of ray-shooting region\\ 
We further reduce the calculation time by minimizing the area of 
the ray-shooting region.  We set the regions of ray shooting in 
the image plane only for which rays arrive in the region around 
caustics on the source plane where the finite-source is important.  
For example, if the perturbation occurs near the peak of a 
high-magnification map, the perturbation is localized in the central 
region around the binary lens components, where a small caustic 
is located. This region in the source plane corresponds to the annulus 
around the Einstein ring in the image plane \citep{griest98, dong06}.  
Then, by shooting rays only in the localized region in the image 
plane, we minimize the number of rays needed for finite-source 
magnification calculations and thus reduce the calculation time.
In Appendix A, we describe how the region of rayshooting is set.
\item
Semi-analytic approximation\\
Finally, we further speed up computations by limiting numerical 
computation of finite-source magnifications only when the source 
is located very close to the caustic and using simple semi-analytic 
hexadecapole approximations \citep{gould08, pejcha09} in the vicinity 
of the caustic.  For this, we divide the source plane into small grids 
and then register the individual grids by different levels depending 
on the distance from the caustic.  Based on the levels, we apply 
different levels of approximations for the magnification calculations.
\end{enumerate}

\begin{table*}
\caption{Fit Parameters. 
${\rm HJD}'={\rm HJD}-2450000$. \label{table:one}}
\begin{tabular}{lccc}
\hline
\hline
\multicolumn{1}{c}{} &
\multicolumn{1}{c}{no} & 
\multicolumn{1}{c}{parallax} & 
\multicolumn{1}{c}{parallax} \\
\multicolumn{1}{c}{} &
\multicolumn{1}{c}{parallax} & 
\multicolumn{1}{c}{($u_0>0$)} & 
\multicolumn{1}{c}{($u_0<0$)} \\
\hline
$\chi^2/{\rm dof}$  & 4067.18/2365                  & 2371.12/2363                 & 2370.08/2363          \\ 
$s$                 & 1.860$\pm$0.015               & 1.705$\pm$0.020              & 1.645$\pm$0.022       \\ 
$q$                 & 0.260$\pm$0.004               & 0.304$\pm$0.008              & 0.295$\pm$0.012       \\
$\alpha$ (deg)      & 106.87$\pm$0.15               & 102.50$\pm$0.29              & -102.34$\pm$0.40      \\
$t_0$ (HJD')        & 4915.48$\pm$0.06              & 4913.72$\pm$0.10             & 4913.85$\pm$0.19      \\
$u_0$               & 0.094$\pm$0.003               & 0.116$\pm$0.004              & -0.132$\pm$0.006      \\
$t_{\rm E}$ (days)  & 153.556$\pm$2.622             & 100.87$\pm$2.315             & 92.49$\pm$2.61        \\
$\rho_\star$        & $(2.9\pm0.3)\times 10^{-4}$   & $(8.4\pm0.6)\times 10^{-4}$  & $(9.5\pm0.7)\times 10^{-4}$  \\
$\pi_{{\rm E},N}$   & --                            & 0.081$\pm$0.015              & -0.120$\pm$0.033      \\
$\pi_{{\rm E},E}$   & --                            & 0.231$\pm$0.008              & 0.252$\pm$0.111       \\
\hline
\end{tabular}
\end{table*}

\begin{figure}
\begin{center}
\includegraphics[width=0.48\textwidth]{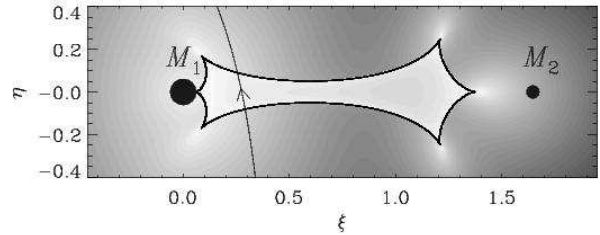}
\end{center}
\caption{
Geometry of the lens system for the best-fit parallax model. The 
two dots represent the locations of the binary lens components
and the closed figure composed of concave curves represents the 
caustic.  The coordinates $(\xi,\eta)$ are centered at the primary 
lens and the abscissa is aligned with the binary axis.  The line 
with an arrow represents the source trajectory.  Note that the 
source trajectory is curved due to the parallax effect. All lengths 
are normalized by the Einstein radius corresponding to the total 
mass of the binary.
}
\label{fig:two}
\end{figure}

\section{Best-fit Model}

From modeling, it is found that OGLE-2009-BLG-023/MOA-2009-BLG-028
is produced by the crossings of a Galactic bulge F-type main-sequence 
source star over the caustic produced by a disk binary lens composed 
of M-type main-sequence stars.  The determined values of the 
normalized star-planet separation and planet/star mass ratio are
\begin{equation}
s=1.65\pm 0.02, \qquad q=0.30\pm 0.01, 
\label{eq5}
\end{equation}
respectively.
In Table \ref{table:one}, we present the lensing parameters 
determined from modeling.  The model light curve is presented in 
Figure \ref{fig:one}.  Figure \ref{fig:two} shows the geometry of 
the lens system under the best-fit model, where the two dots represent 
the locations of the lens components, the closed figure composed of 
concave curves represents the caustic produced by the binary lens, 
and the line with an arrow represents the source trajectory.  We 
note that the trajectory is curved due to the parallax effect.  
We find no serious local minima except the one caused by the 
mirror-image symmetry between the source trajectories with the 
impact parameters and orientation angles of the source trajectory 
of $(u_0,\alpha)$ and $(-u_0,-\alpha)$.  We find that the solution 
with $u_0<0$ is preferred over the solution with $u_0>0$ by 
$\Delta\chi^2=1.0$, implying that it is difficult to distinguish 
the two solutions.

\begin{figure}
\begin{center}
\includegraphics[width=0.48\textwidth]{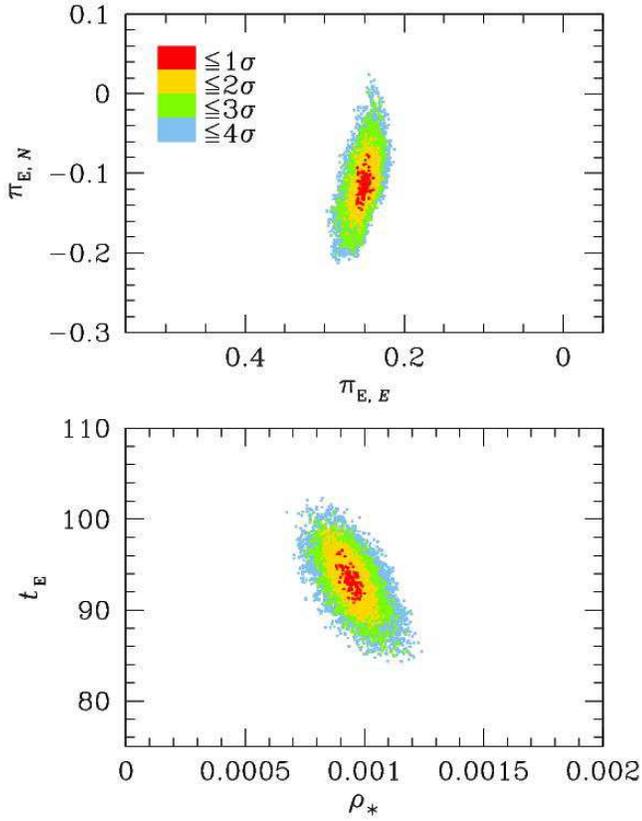}
\end{center}
\caption{
Upper panel: 
Contour of $\chi^2$ in the parameter space of $(\pi_{{\rm E},E},
\pi_{{\rm E},N})$, which are the components of the lens-parallax
vector $\pivec_{\rm E}$ projected in the sky in the north and east 
celestial coordinates, respectively.
Lower panel: 
Contours of $\chi^2$ in the parameter space of the normalized 
source radius $\rho_\star$ and the Einstein time scale $t_{\rm E}$.
}
\label{fig:three}
\end{figure}

\subsection{Lens Parallax}

Light curves of long time-scale events are susceptible to the parallax 
effect.  The time scale of the event OGLE-2009-BLG-023/MOA-2009-BLG-028 
is $t_{\rm E}\sim 100\ {\rm days}$, which comprises nearly 1/3 of the 
orbital period of the Earth.  We, therefore, search for solutions 
considering the parallax effect.  From this, we find that the parallax 
model provides a significantly better fit with $\Delta\chi^2=1697$.  
The determined values of the parallax parameters are 
\begin{equation}
\pi_{{\rm E},N}=-0.12\pm 0.03, \qquad \pi_{{\rm E},E}=0.25\pm 0.01.
\label{eq6}
\end{equation}
The upper panel of Figure \ref{fig:three} shows the scatter plot of 
$\chi^2$ in the parameter space of $(\pi_{{\rm E},E}, \pi_{{\rm E},N})$.  
The contours are elongated along the $\pi_{{\rm E},N}$ axis because 
the apparent motion of the Sun at $t_0$ projected onto the sky is 
perpendicular to the $\pi_{{\rm E},N}$ axis.

It is known that the orbital motion of a binary source can give 
rise to distortions of lensing light curves (`xallarap' effect) that 
are similar to those induced by the parallax effect \citep{smith03}.  
We check this possibility by conducting xallarap modeling under the 
assumption that the binary source is in a circular orbit.  From this, 
we find that the improvement of the fit of the xallarap model from
that of the parallax model is merely $\Delta\chi^2=2.4$, which is much 
smaller than the improvement by the parallax effect of $\Delta\chi^2
\sim 1700$.  In addition, the minimum $\chi^2$ occurs at an orbital 
period of $P\sim 1$ yr, which corresponds to the orbital period of 
the Earth around the Sun.  Both facts support the parallax 
interpretation of the light curve deviation.

\begin{figure}
\begin{center}
\includegraphics[width=0.48\textwidth]{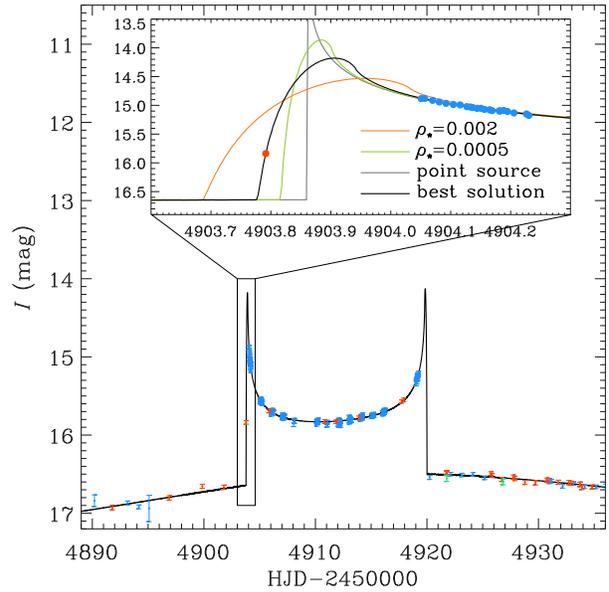}
\end{center}
\caption{
Enlargement of the caustic-crossing part of the light curve.
The inset shows the light curve during the caustic entrance of 
the source star.  For comparison, we draw light curves for 
different values of the normalized source radius $\rho_\ast$.
}
\label{fig:four}
\end{figure}

\begin{table}
\caption{Physical Parameters \label{table:two}}
\begin{tabular}{lcc}
\hline
\hline
\multicolumn{1}{c}{} &
\multicolumn{1}{c}{parallax ($u_0>0$)} & 
\multicolumn{1}{c}{parallax ($u_0<0$)} \\
\hline
$\theta_{\rm E}$ (mas)           & 1.583$\pm$0.191  &  1.475$\pm$0.187  \\
$\mu$            (mas yr$^{-1}$) & 5.727$\pm$0.692  &  5.822$\pm$0.738  \\
$D_{\rm L}$      (kpc)           & 1.933$\pm$0.203  &  1.845$\pm$0.214  \\ 
$M_1$            ($M_{\odot}$)   & 0.608$\pm$0.077  &  0.500$\pm$0.071  \\
$M_2$            ($M_{\odot}$)   & 0.185$\pm$0.023  &  0.148$\pm$0.021  \\
\hline
\end{tabular}
\end{table}

\subsection{Einstein Radius}

The Einstein radius is measured from the normalized source radius 
$\rho_\star$ combined with the information about the angular source 
radius $\theta_\star$ by
\begin{equation}
\theta_{\rm E}={\theta_\star \over \rho_\star}.
\label{eq7}
\end{equation}
The normalized source radius is measured with a moderate uncertainty 
from the analysis of the light curve during the caustic crossings.  
In the lower panel of Figure \ref{fig:four}, we present the enlargement 
of the caustic-crossing part of the light curve.  From the figure, it 
is found that a single OGLE data point taken during the time when the 
source is on the fold caustic (at ${\rm HJD} \sim 4903.78$) and 
multiple MOA data points taken at the time when the source is about 
to leave the caustic (during $4904.05 \lesssim {\rm HJD} \lesssim 
4904.20$) provide constraints on $\rho_\star$.  The measured value 
of the normalized source radius is 
\begin{equation}
\rho_\star = (9.5 \pm 0.7) \times 10^{-4}.
\label{eq8}
\end{equation}
Considering that the normalized source radius of a main-sequence source 
star is $\rho_\star\sim (O) 10^{-3}$ for typical lensing events caused 
by low-mass lens stars located half way between the source and observer,
the measured value of $\rho_\star$ is substantially small.  This means 
that the Einstein radius is big, suggesting that either the lens is 
heavy or it is located close to the observer.  The lower panel of 
Figure \ref{fig:three} shows the $\chi^2$ distribution in the parameter 
space of $\rho_\star$ and $t_{\rm E}$.

\begin{figure}
\begin{center}
\includegraphics[width=0.48\textwidth]{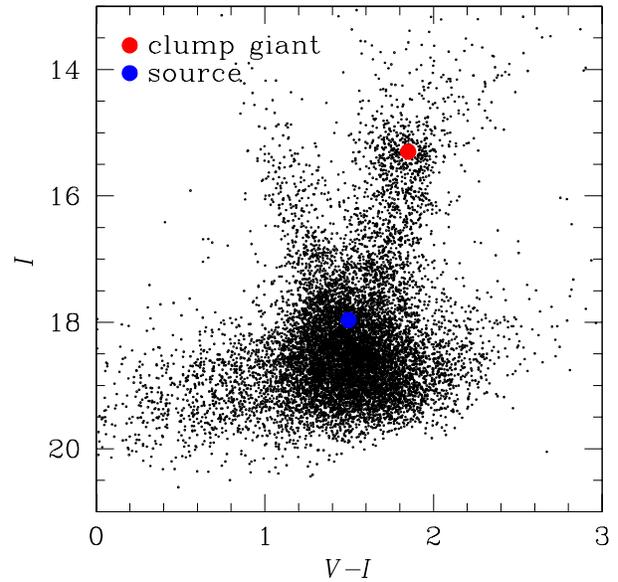}
\end{center}
\caption{
Position of the source star in the instrumental color-magnitude 
diagram constructed based on OGLE data.
}
\label{fig:five}
\end{figure}

The angular source radius is determined from the information of the 
de-reddened color of the source star \citep{yoo04}.  We determine the 
color by using the centroid of clump giant stars in the color-magnitude 
diagram as a reference position \citep{stanek94, stanek97} under the 
assumption that the source star and clump giants experience the same 
amount of extinction.\footnote{For source stars in the bulge, the 
uncertainty caused by this assumption would be small. This is because 
the extinction toward the Galactic bulge field is caused by the dust 
distributed in the disk and thus even if there exist some differences 
between the distances to the clump centroid and the source star, the 
amount of extinction would be nearly the same.  The source star can 
be located in the disk but the probability is low. The validity of 
the source location in the bulge can also be checked by its location 
on the color-magnitude diagram. From the diagram presented in Figure 
\ref{fig:five}, it is found that the source is located in the region 
where bulge main-sequence stars are densely populated, suggesting that 
the source is located in the bulge.}
The offset between the source and clump centroid is measured in the 
instrumental color-magnitude diagram that is constructed by using the 
OGLE $V$ and $I$ band images taken toward the bulge field where the 
source is located (Figure~\ref{fig:five}).  Then, the angular source size 
is determined by first transforming from $(V-I)_0$ to $(V-K)_0$ using 
the color-color relation of \citet{bessel98} and then applying the 
relation between $(V-K)_0$ and the angular stellar radius of 
\citet{kervella04}.  For the best-fit model, we find that the 
de-reddened magnitude and color of the source star are $I_0=16.7$ and 
$(V-I)_0=0.67$, respectively, implying that the source is an F-type 
main-sequence star with an angular radius of
\begin{equation}
\theta_\star=(1.40\pm 0.15)\ \mu{\rm as}.
\label{eq9}
\end{equation}
Here we adopt an average distance to clump giants toward the field 
of 7.7 kpc estimated based on the Galactic mass distribution model 
of \citet{han03}.  The uncertainty  of $\theta_\star$ is determined 
from the combination of the uncertainty of the distance to the source 
and an additional $7\%$ intrinsic error in the conversion process 
from the measured color to the source radius.

With the measured values of the normalized and angular source radii, 
the Einstein radius is determined as
\begin{equation}
\theta_{\rm E} = (1.48\pm 0.19)\ {\rm mas}.
\label{eq10}
\end{equation}
This corresponds to the relative lens-source proper motion of 
\begin{equation}
\mu = {\theta_{\rm E}\over t_{\rm E}}
=(5.82 \pm 0.74)\ {\rm mas}\ {\rm yr}^{-1}.
\label{eq11}
\end{equation}

\subsection{Physical Parameters}

With the measured Einstein radius and parallax, the mass and distance 
to the lens are determined from the relations in equation (\ref{eq4}). 
For the best-fit model, these values are 
\begin{equation}
M=(0.65 \pm 0.09) \ \ M_\odot
\label{eq12}
\end{equation}
and
\begin{equation}
D_{\rm L}=(1.8 \pm 0.2) \ \ {\rm kpc},
\label{eq13}
\end{equation}
respectively.  With the determined mass ratio, it is found that the 
masses of the individual lens components are $M_1=(0.50\pm 0.07)\ 
M_\odot$ and $M_2=(0.15\pm 0.02)\ M_\odot$, respectively.  Therefore, 
the lens is composed of an early and a late M-type main-sequence stars 
located in the Galactic disk.  The model with $u_0>0$ yields the distance 
and mass of the lens slightly bigger than those of the model with $u_0<0$ 
(see Table \ref{fig:two}).

\section{Discussion and Conclusion}

We analyzed the light curve of the long time-scale, caustic-crossing, 
binary-lens event OGLE-2009-BLG-023/MOA-2009-BLG-028.  Despite that 
the event was observed solely by survey experiments, we could 
uniquely determine the mass of the lens and distance to it by 
simultaneously measuring the Einstein radius and lens parallax.  
It was turned out that the event was produced by the crossings of 
a Galactic bulge F-type main-sequence source star over the caustic 
produced by a disk binary lens composed of M-type main-sequence 
stars.  The event demonstrates that the physical parameters of 
binary lenses can be uniquely determined from data obtained by 
survey observations.

Microlensing can potentially probe the distributions of binaries as
functions of mass ratio and separation, that can provide important
observational constraints on theories of star formation \citep{gould01}. 
Especially, microlensing is sensitive to low-mass companions that are 
difficult to be detected by other methods and thus it is possible to
make complete distributions down to lower mass limit of binary
companions.   Due to this importance, there have been several 
systematic searches for binary lenses \citep{alcock00,jaroszynski02, 
jaroszynski04, jaroszynski05, jaroszynski06, skowron07}.  However, 
the samples of binaries acquired from previous surveys were not
adequate enough to strongly constrain the binary distributions.
One important reason for this is the difficulty in estimating the 
detection efficiency of binary-lens events. Most binary-lens events 
are detected through the channel of caustic-crossing events where 
caustic crossings were accidently discovered by the sudden rise of 
the source star flux.  Due to this haphazard nature of binary-lens 
events, it is difficult to estimate the detection efficiency that 
is essential for the statistical studies of binaries. In addition, 
the physical quantities of lenses for most of binary-lens events 
could not be determined, making detailed studies of binaries difficult.

However, the situation will be different with the advent of 
new-generation experiments. As mentioned, the recent upgrades of 
the OGLE and MOA experiments already significantly increased the 
observational cadence of the surveys. In addition to these experiments, 
there is a planned experiment that can increase the cadence even 
higher.  Korea Microlensing Telescope Network (KMTNet) is an approved 
project that will employ three telescopes, each of which will have 
a 1.6 m aperture and 4 deg$^2$ field of view. They will be located 
in three different continents of South America (Chile), Africa 
(South Africa), and Australia for continuous observations of 
microlensing events.  The expected cadence of the experiment is 
6 hr$^{-1}$.  Considering that the typical time scale of caustic 
crossings is several hours for events involved with main-sequence 
stars and extends to $\gtrsim 10$ hrs for events associated with 
giant source stars,  the cadence of the survey is high enough to 
resolve caustic crossings of most events, enabling measurements 
of Einstein radii.  For a significant fraction of these events, 
it will be possible to additionally measure lens parallaxes, 
enabling to completely determine the physical parameters of binary 
lenses.  Furthermore, the uniform coverage of events will make it 
easier to estimate the detection efficiency of binary-lens events, 
enabling statistical analysis of binary distributions possible.  
Therefore, future lens surveys can provide a new way to study 
Galactic binary stars.

\section*{Acknowledgements}

We acknowledge the following support:
National Research Foundation of Korea 2009-0081561 (CH);
Polish MNiSW N20303032/4275 (AU); 
MEXT19015005, JSPS18749004 (TS); 
Grants JSPS18253002, JSPS20340052, and JSPS19340058 (MOA);
NSF AST-0708890, NASA NNX07AL71G (DPB);
Marden Fund of NZ (IAB, JBH, DJS, SLS, PCMY);
Foundation for Research Science and Technology of NZ (IAB);
NSF AST-0757888 (AG).

\appendix
\section{Region of Rayshooting}

We determine the range of rayshootinh as follows.
For sources close to a fold caustic, the point-source magnification 
is represented by 
$$
A_{\rm p}=A_0 + (d/u_\perp)^{-1/2}\Theta(d),
$$
where $d$ is the perpendicular distance to the caustic with $d>0$ for 
sources interior to the caustic, $u_\perp$ represents the characteristic 
strength of the local caustic, $\Theta(x)$ is the Heaviside step function, 
and $A_0$ represents the total magnification of the slowly varying images 
\citep{schneider86, gaudi02}.  Then, under the approximation that the
magnification variation of the slowly varying images is negligible,
the point-source approximation can be used for magnifications in the 
region outside the caustic with $|d|>\rho_\star$.  Inside the caustic, 
the finite-source magnification, $A_{\rm f}$, is expressed as 
$$
A_{\rm f}\propto \rho_\star^{-2} \int_{-\rho_\star}^{\rho_\star}
\left( {\rho_{\star}^2-r^2 \over d+r }\right)^{1/2} dr=
d^{-1/2} f(d/\rho_\star).
$$
Then, the fractional deviation from the point-source magnification is 
$$
\epsilon = {A_{\rm f}-A_{\rm p}\over A_{\rm p}}
=\epsilon (d/\rho_\star),
$$
implying that the fraction deviation depends only on the ratio 
$d/\rho_\star$.  We find that the deviation are 2\%, 1\%, and 0.5\% 
for the distances from the caustic of $d=2.3\rho_\star$, $3.1\rho_\star$,
and , $4.3\rho_\star$, respectively, implying that the deviation 
decreases rapidly with the increase of 
$d/\rho_\star$.  
To be conservative, we set the range of rayshooting as 
$d<2\rho_\star$ and $d<5\rho_\star$ 
inside and outside the caustic, respectively. 

\end{document}